\documentclass[12pt,a4paper]{conference}
\usepackage{fancyhdr}
\usepackage{graphicx,amsmath,amssymb,cite}
\usepackage{multind}
\makeindex{author} \makeindex{subject}
\newcommand{\beq}{\begin{equation}}
\newcommand{\eeq}[1]{\label{#1}\end{equation}}
\newcommand{\eeqn}{\end{equation}}

\newcommand{\beqa}{\begin{eqnarray}}
\newcommand{\eeqa}[1]{\label{#1}\end{eqnarray}}
\newcommand{\eeqan}{\end{eqnarray}}

\def\Journal#1#2#3#4{{#1} {\bf #2}, #3 (#4)}

\newcommand{\NPB}{Nucl. Phys. B}
\newcommand{\PLB}{Phys. Lett.  B}
\newcommand{\PRL}{Phys. Rev. Lett.}
\newcommand{\PRD}{Phys. Rev. D}

\let\bar=\overbar

\newcommand{\Dslash}{\not{\hbox{\kern-4pt $D$}}}
\newcommand{\dslash}{\not{\hbox{\kern-2pt $\del$}}}

\newcommand{\msb}{{\bar{\ssstyle M \kern -1pt S}}}

\def\Journal#1#2#3#4{{#1}~{\bf #2}, #3 (#4)}  

\def\NIMA{{\em Nucl.~Instrum.~Methods}~A}

\def\NPB{{\em Nucl.~Phys.}~B}
\def\PLB{{\em Phys.~Lett.}~B}
\def\PRL{\em Phys.~Rev.~Lett.}
\def\PRD{{\em Phys.~Rev.}~D}

\def\EPJ{{\em Eur.~Phys.~J.}~C}
\def\JPG{{\em J.~Phys.}~G}

\newcommand{\kl}{K_L}

\newcommand{\klpipipi}{\kl \to \pi^0 \pi^0 \pi^0}

\newcommand{\klpipipic}{\kl \to \pi^+ \pi^- \pi^0}

\newcommand{\kpmpipipi}{K^\pm \to \pi^\pm \pi^+ \pi^-}
\newcommand{\kppipipi}{K^+ \to \pi^+ \pi^+ \pi^-}

\newcommand{\kpmpipizpiz}{K^\pm \to \pi^\pm \pi^0 \pi^0}
\newcommand{\kppipizpiz}{K^+ \to \pi^+ \pi^0 \pi^0}

\newcommand{\kpmpipie}{K^\pm \to \pi^+ \pi^- e^\pm \nu (\bar{\nu})}

\newcommand{\bdm}{\begin{displaymath}}
\newcommand{\edm}{\end{displaymath}}

\begin{document}


\Chapter{WIGNER-CUSP IN KAON DECAYS AND DETERMINATION OF $\pi\pi$ SCATTERING LENGTHS}
           {Determination of $\pi\pi$ Scattering Lengths in Kaon Decays}{R.~Wanke}
\vspace{-6cm}
\includegraphics[width=6 cm]{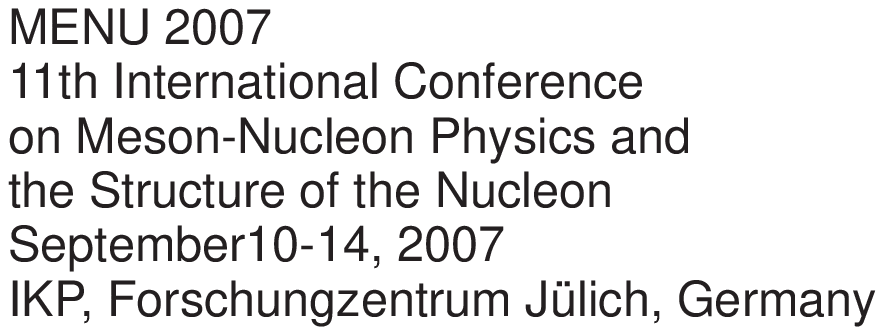}
\vspace{4cm}

\addcontentsline{toc}{chapter}{{\it R.~Wanke}}
\label{authorStart}


\begin{raggedright}

{\it R.~Wanke}\index{author}{Wanke, R.}\\
Institut f\"ur Physik\\
Universit\"at Mainz\\
D-55099 Mainz, Germany
\bigskip\bigskip

\end{raggedright}

\begin{center}
\textbf{Abstract}
\end{center}
In the last few years it has become possible to study low energy $\pi\pi$ scattering
in kaon decays to three pions, thanks to the high statistics measurement of 
$\kpmpipizpiz$ decays performed by the NA48/2 experiment at the CERN SPS. 
At the $\pi^+ \pi^-$ threshold, the $\pi^0 \pi^0$ mass spectrum exhibits a Wigner-cusp,
from which the $S$-wave $\pi\pi$  scattering lengths are extracted with high precision.
This measurement is complementary to the extraction of the scattering lengths from 
$K_{e4}$ decays, which is also performed by the NA48/2 experiment.

\section{Introduction}

In the low-energy regime, the perturbative description of the strong interaction 
breaks down, as the strong coupling constant becomes of ${\cal O}(1)$. 
Chiral Perturbation Theory (ChPT) is an effective theory, which circumvents this problem
by making use of the chiral symmetry of the theory in the limit of 
vanishing quark masses.
Spontaneous breaking of the chiral symmetry generates 8 pseudo-scalar
Goldstone bosons, among them pions and kaons. They obtain their small but non-zero 
masses by the additional symmetry breaking of non-vanishing quark masses.
In the framework of ChPT, the values of the iso-spin 0 and 2 $S$-wave $\pi\pi$ scattering lengths $a_0$ and $a_2$
are directly connected with the size of the chiral condensate and the pion mass~\cite{bib:gasserleutwyler}.
The scattering lengths can accurately be predicted
to $a_0 m_{\pi^+} = 0.220 \pm 0.005$ and $a_2 m_{\pi^+} = -0.044 \pm 0.010$~\cite{bib:cgl}.
Thus, precise measurements of the $\pi\pi$ scattering lengths are a crucial test of ChPT.

Previous measurements have traditionally been performed in the 
semileptonic decay $\kpmpipie$ ($K_{e4}$). An early measurement
by the Geneva-Saclay experiment analyzed 30000 events~\cite{bib:gesaclay}. 
More recently, the BNL experiment E865 has measured $(a_0-a_2) m_{\pi^+} = 0.258 \pm 0.013$
from about 400000 $K_{e4}$ events~\cite{bib:e865}.
Another recent determination of the scattering lengths has been carried out by the
DIRAC experiment at CERN from the lifetime of pionium atoms. They obtain
$(a_0-a_2) m_{\pi^+} = 0.264 {+0.020 \atop -0.011}$ from an analysis 
of a part of their data~\cite{bib:dirac}.

Here, new measurements of the NA48/2 Collaboration are reported.
In $\kpmpipizpiz$ decays, a Wigner-cusp from $\pi\pi$ rescattering in the decay amplitude
was discovered at $m(\pi^0 \pi^0) = 2 \, m_{\pi^+}$.
This allows a very precise determination of $a_0$ and $a_2$ with a completely new method.
In addition, the NA48/2 Collaboration has also determined $a_0$ and $a_2$ in $K_{e4}$ decays, 
with a greatly improved precision with respect to previous measurements.

\begin{figure}[t]
  \begin{center}
    \includegraphics[width=\linewidth]{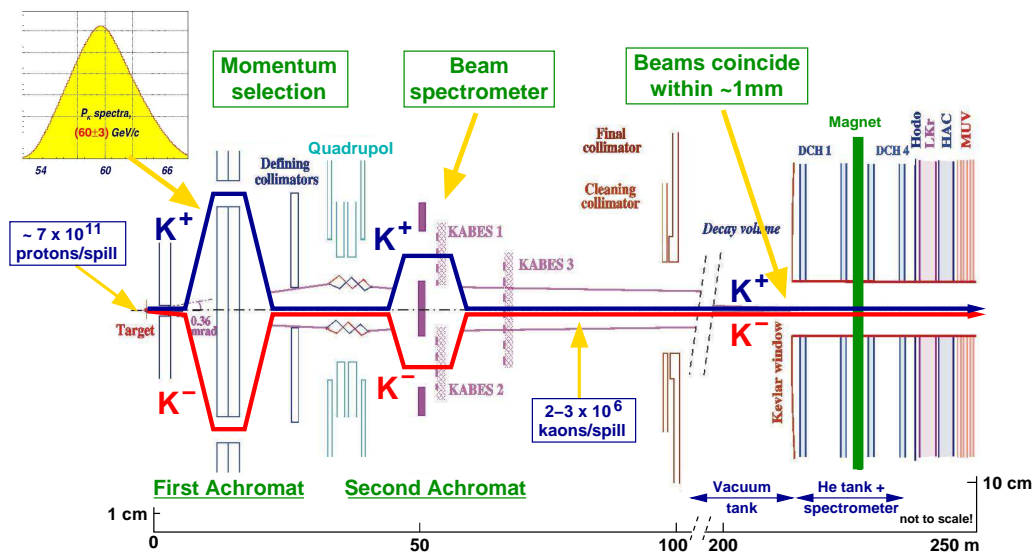}
    \caption{Set-up of the NA48/2 experiment.} 
    \label{fig:na48_2}
  \end{center}
\end{figure}

The NA48/2 experiment has been taking data in the years 2003 and 2004.
From a 400~GeV/$c$ proton beam, positive and negative kaons 
with a momentum bite of $p_K = (60 \pm 3)$~GeV/$c$ were simultaneously
selected by a system of achromats (see Fig.~\ref{fig:na48_2}). 
A detailed detector description can be found in~\cite{bib:na48det}.
The main aim of the experiment was the search for direct CP violation in decays
of charged kaons into three pions.
The trigger therefore was designed to efficiently 
select events with three charged tracks as well as $\kpmpipizpiz$ events.
In total, about $2\times10^9$ three-track events and about $90\times10^6$~$\kpmpipizpiz$ 
events were recorded in the two years of data-taking.

\section{Wigner-Cusp and $\pi\pi$ Rescattering in\\ $\kpmpipizpiz$ Decays}

\begin{figure}[t]
  \begin{center}
    \includegraphics[width=0.45\linewidth]{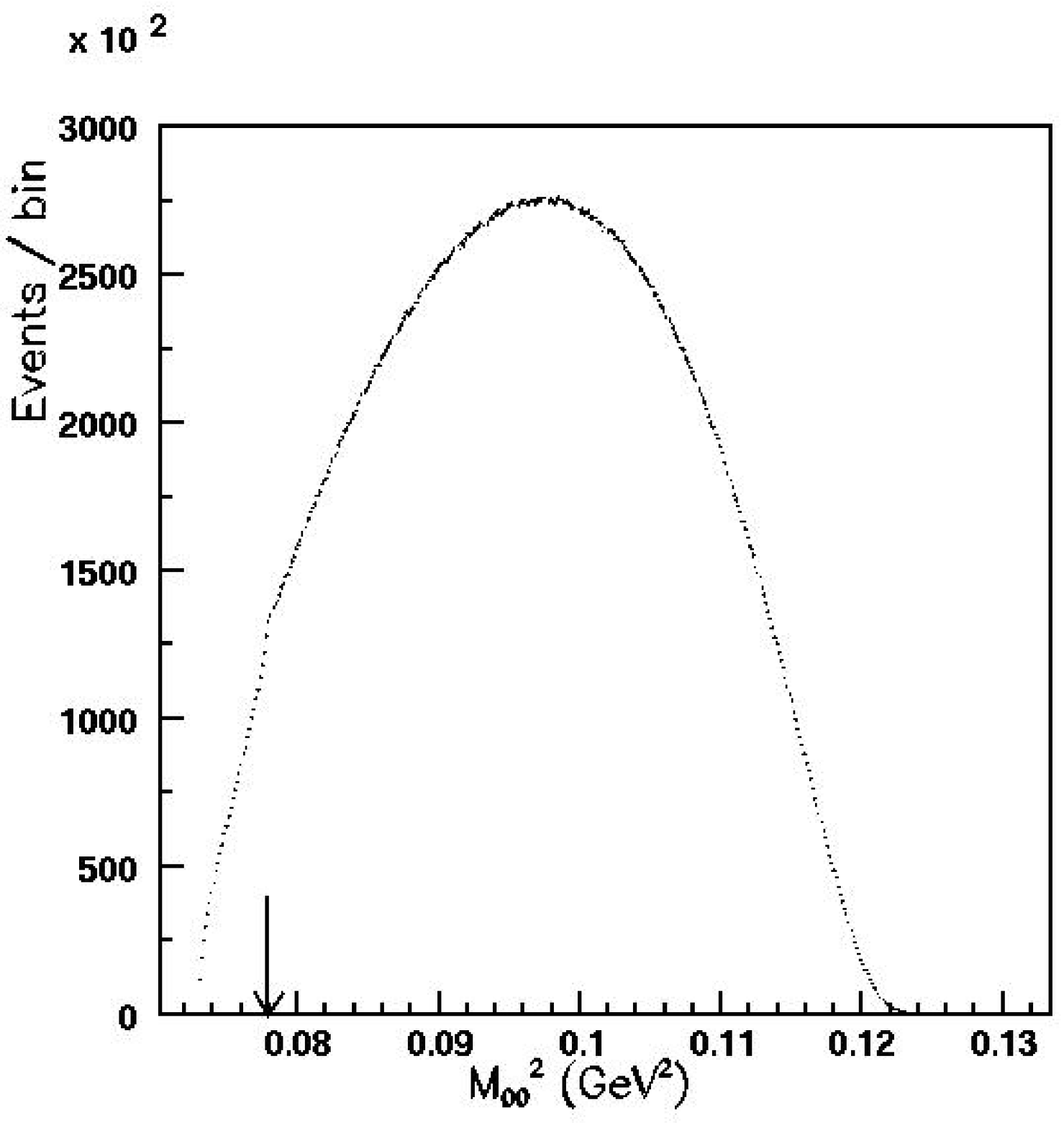}
    \hspace{\fill}
    \includegraphics[width=0.45\linewidth]{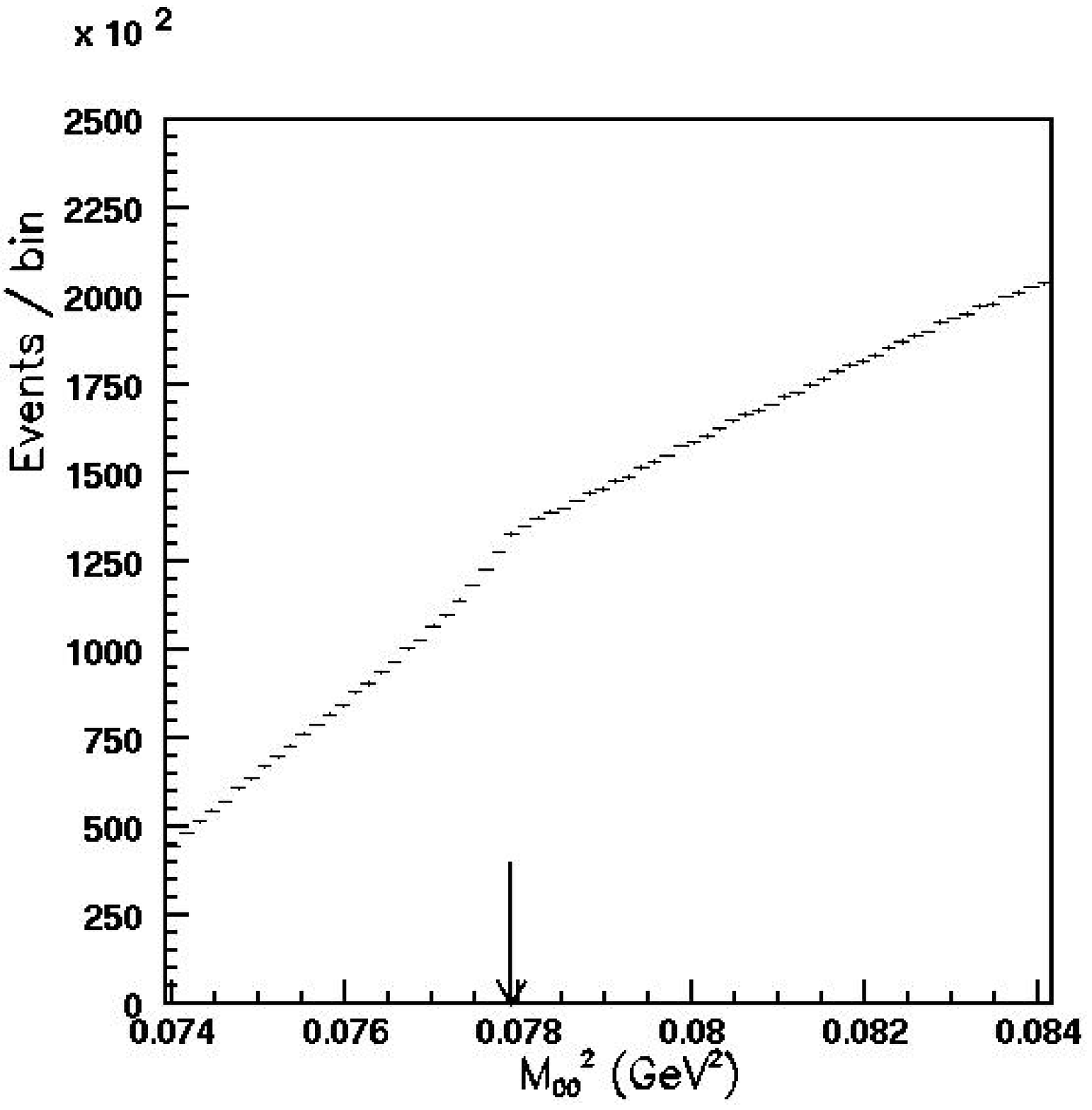}
    \caption{Distribution of $m^2_{\pi^0\pi^0}$ of the NA48/2 $\kpmpipizpiz$ data.
             Left: Full kinematic region. Right: Zoom into the cusp region. The arrow marks
             the $(2 \, m_{\pi^+})^2$ threshold.}
    \label{fig:na48cusp}
  \end{center}
\end{figure}

With the unprecedented statistics collected in the channel $\kpmpipizpiz$,
it was possible for NA48/2 to precisely measure the distribution of the
$\pi^0 \pi^0$ invariant mass. The original aim of this analysis was the search
for the decay $K^\pm \to \pi^\pm (\pi^+ \pi^-)_\text{atom}$ with the subsequent decay
of the pionium atom $(\pi^+ \pi^-)_\text{atom}$ into $\pi^0 \pi^0$.
It came as a surprise, that the $\pi^0 \pi^0$ invariant mass distribution exhibited
a distinct and clearly visible cusp at the $\pi^+ \pi^-$ mass threshold (see Figure~\ref{fig:na48cusp}).

After its discovery, the effect was explained by N.~Cabibbo~\cite{bib:cabibbo} as Wigner cusp,
arising from $\pi^+ \pi^- \to \pi^0 \pi^0$ rescattering from the decay $\kpmpipipi$~\footnote{Independently, 
the effect had already been predicted in $\pi \pi$ scattering much earlier by 
U.-G.~Mei{\ss}ner and collaborators~\cite{bib:meissner}.}.
As consequence, a rescattering amplitude ${\cal M}_1$ from $\kpmpipipi$ has to be added to the unperturbed
$\kpmpipizpiz$ amplitude ${\cal M}_0$. Below the cusp point  $m_{\pi^0 \pi^0} = 2 \, m_{\pi^+}$,
${\cal M}_1$ is real and negative, thus explaining the observed deficit of events.
In the one-loop approximation, ${\cal M}_1$ turns imaginary above the cusp. In this region, 
the effect on $m_{\pi^0 \pi^0}$ is much smaller and less visible.
The cusp strength is proportional to $(a_0-a_2) m_{\pi^+}$, the
effect therefore allows the extraction of the $\pi\pi$ scattering lengths.

\begin{figure}[t]
  \begin{center}
    \includegraphics[width=0.32\linewidth]{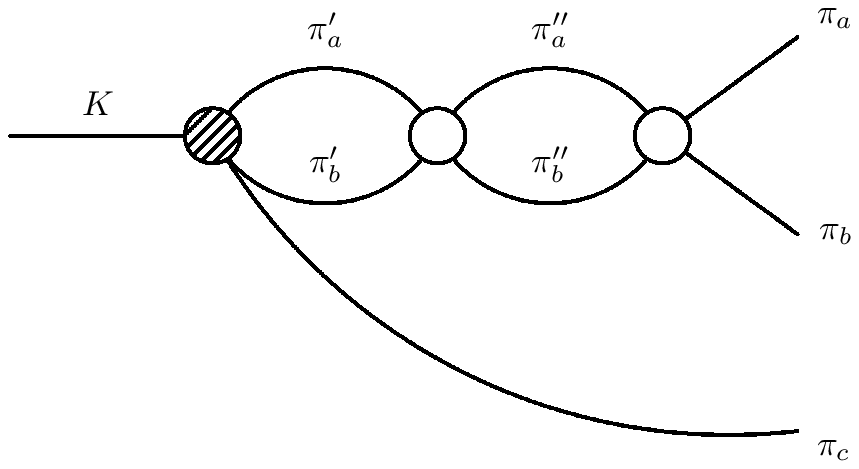}
    \includegraphics[width=0.32\linewidth]{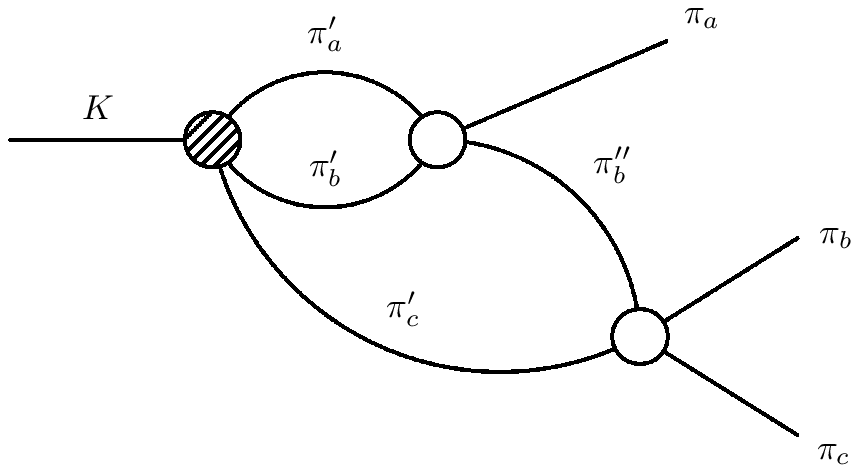}
    \includegraphics[width=0.32\linewidth]{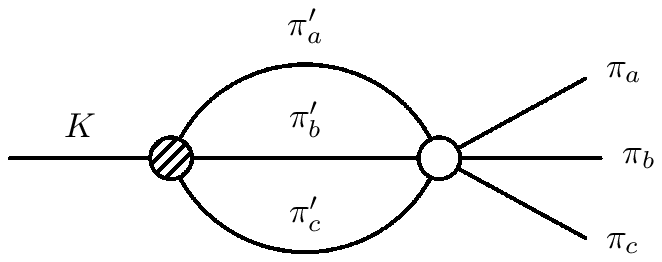}
    \caption{Diagrams of two-loop ${\cal O}(a_i^2)$ corrections (from ref.~\cite{bib:cabibboisidori}).} 
    \label{fig:rescatteringdiagrams}
  \end{center}
\end{figure}

Second order calculations were performed by Cabibbo and Isidori~\cite{bib:cabibboisidori}.
These include two-loop level ${\cal O}(a_i^2)$ corrections (Fig.~\ref{fig:rescatteringdiagrams}),
which allow the separate determination of the scattering length $a_2$.
Also other rescattering processes as $\pi^0 \pi^0 \to \pi^0 \pi^0$, $\pi^+ \pi^0 \to \pi^+ \pi^0$, etc. are considered,
and all $K\to 3 \pi$ decays as e.g.\ $\klpipipi$ are covered.
From this, the theoretical error on the extraction of $a_0 - a_2$ is about $5\%$.
To reach a higher level of precision, ${\cal O}(a_i^3)$ and radiative corrections 
have to be taken into account.

\begin{figure}[t]
  \begin{center}
    \includegraphics[width=0.7\linewidth]{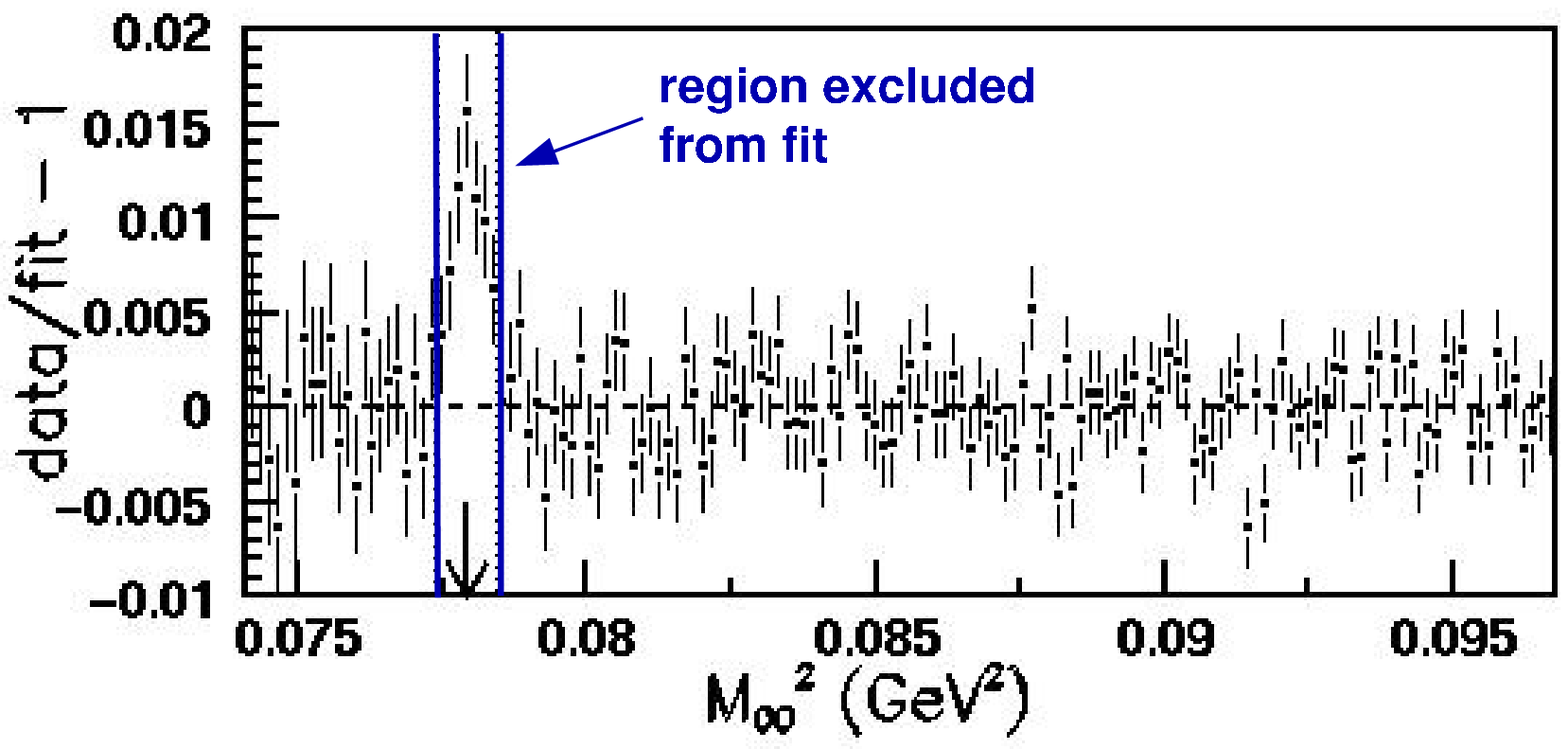}
    \caption{Pulls of the fit to the $m^2_{\pi^0\pi^0}$ distribution.
             The region between the vertical lines was excluded from the fit.}
    \label{fig:pionium}
  \end{center}
\end{figure}

Using the Cabibbo-Isidori model, the NA48/2 collaboration has performed a fit to
the $m^2_{\pi^0 \pi^0}$ distribution seen in $\kpmpipizpiz$~\cite{bib:na48cusp}, 
following a previous analysis on a fraction of the data~\cite{bib:na48cuspold}.
Free fit parameters were the scattering lengths $a_0-a_2$, $a_2$, 
the linear and quadratic slopes $g$ and $h$ of the Dalitz plot 
$u$ distribution, and the normalisation.
The region around the cusp point at $m^2_{\pi^0 \pi^0} = (2 \, m_{\pi^+})^2$ was excluded
from the fit to avoid an influence of possible pionium formation.
The result of the fit is
\begin{eqnarray}
\nonumber  (a_0 -a_2) \, m_{\pi^+} & = &  0.261 \pm 0.006_\text{stat} \pm 0.003_\text{syst} \pm 0.001_\text{ext} \pm 0.013_\text{theory} \\
                  a_2 \, m_{\pi^+} & = & -0.037 \pm 0.013_\text{stat} \pm 0.009_\text{syst} \pm 0.002_\text{ext}.
\end{eqnarray}
The result is still preliminary.
The residuals of the fit are shown in Fig.~\ref{fig:pionium}. 
When the correlation of $\rho = -0.92$ between both values are taken into account, 
the result on the scattering length $a_0$ is
\begin{equation}
  a_0 \, m_{\pi^+} = 0.224 \pm 0.008_\text{stat} \pm 0.006_\text{syst} \pm 0.003_\text{ext}\pm 0.013_\text{theory}.
\end{equation}
The main contributions to the systematic uncertainty come from the shower simulation
in the calorimeter, the trigger efficiency, 
and the dependence of the Dalitz plot $v$ parameter. 
The external error comprises the ratio between the $\kppipipi$ and $\kppipizpiz$ 
decay widths~\cite{bib:pdg06}. The theoretical uncertainty is estimated due
to the neglected ${\cal O}(a_i^3)$ and radiative corrections, as mentioned above.

Using analyticity and chiral symmetry, $a_0$ and $a_2$ are directly related with each other~\cite{bib:cgl2}.
With this constraint, the experimental error is reduced significantly and the 
NA48/2 result becomes 
$(a_0 -a_2) \, m_{\pi^+} =  0.263 \pm 0.003_\text{stat} \pm 0.001_\text{syst} \pm 0.001_\text{ext} \pm 0.013_\text{theory}$.

When including the region around the cusp into the fit and allowing for
an additional pionium component in the fit, NA48/2 observed
an excess of $(1.8 \pm 0.2) \times 10^{-5}$ events per $\kpmpipipi$ decay
at the cusp point (see Fig.~\ref{fig:pionium}).
This is a factor of $2.3 \pm 0.3$ larger than the predicted rate
of $0.8 \times 10^{-5}$ for pionium formation in $\kpmpipipi$~\cite{bib:silagadze}.
However, it has recently pointed out~\cite{bib:gevorkian},
that electromagnetic corrections to final state interactions have to be taken
into account, which might explain the observed excess over the prediction.

\section{Rescattering in $\klpipipi$ decays}

In $\klpipipi$ events, a similar cusp at $m_{\pi^0 \pi^0} = 2 \, m_{\pi^+}$
is expected~\cite{bib:bissegger}. By computing the matrix elements for
$K^+/K_L \to 3\pi$ at the cusp point from the measured partial widths and
Dalitz plot slopes~\cite{bib:pdg06}, the visibility
is expected to be about 13 times smaller in $\klpipipi$ compared to
$\kpmpipizpiz$.

\begin{figure}[t]
  \begin{center}
    \includegraphics[width=0.39\linewidth]{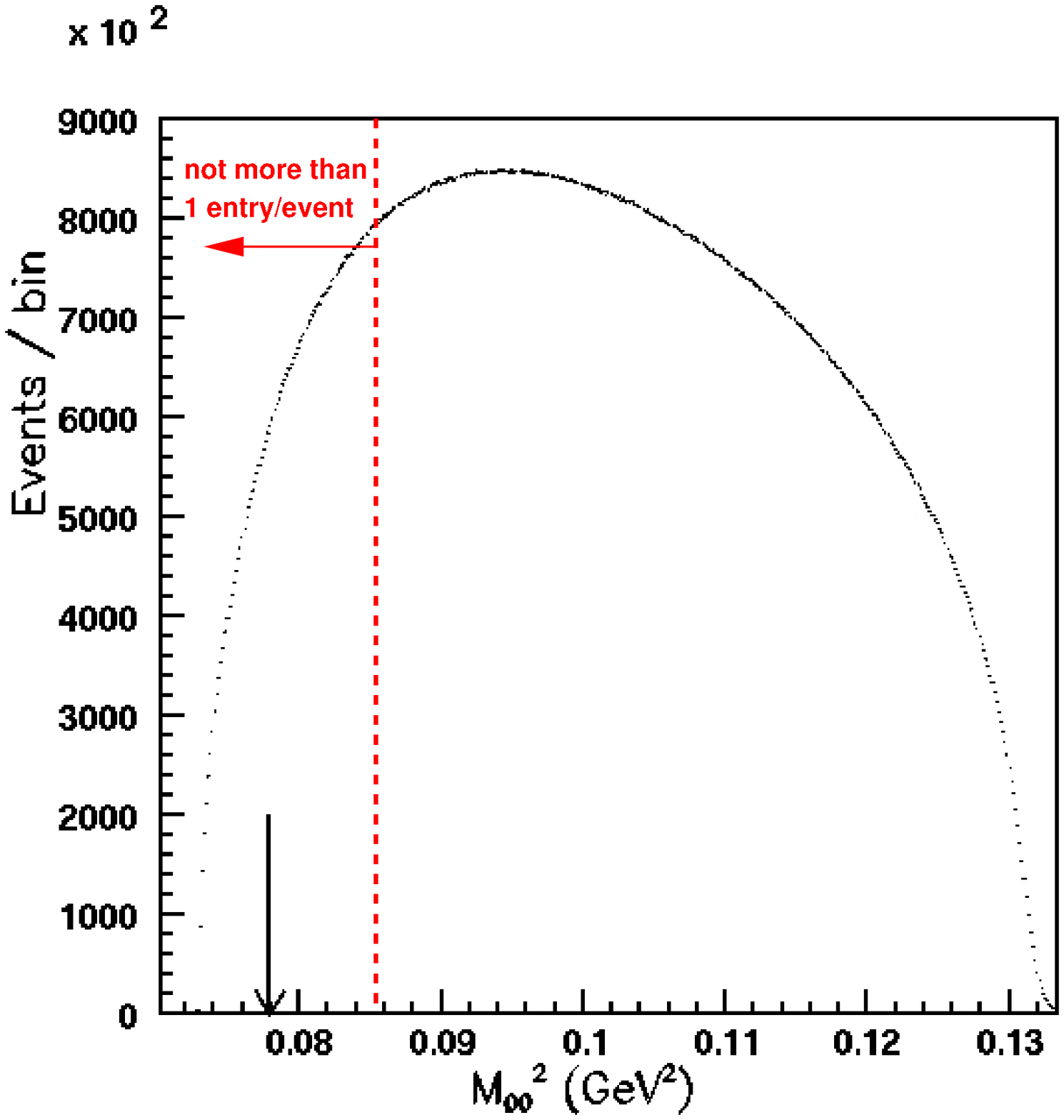}
    \hspace{\fill}
    \includegraphics[width=0.58\linewidth]{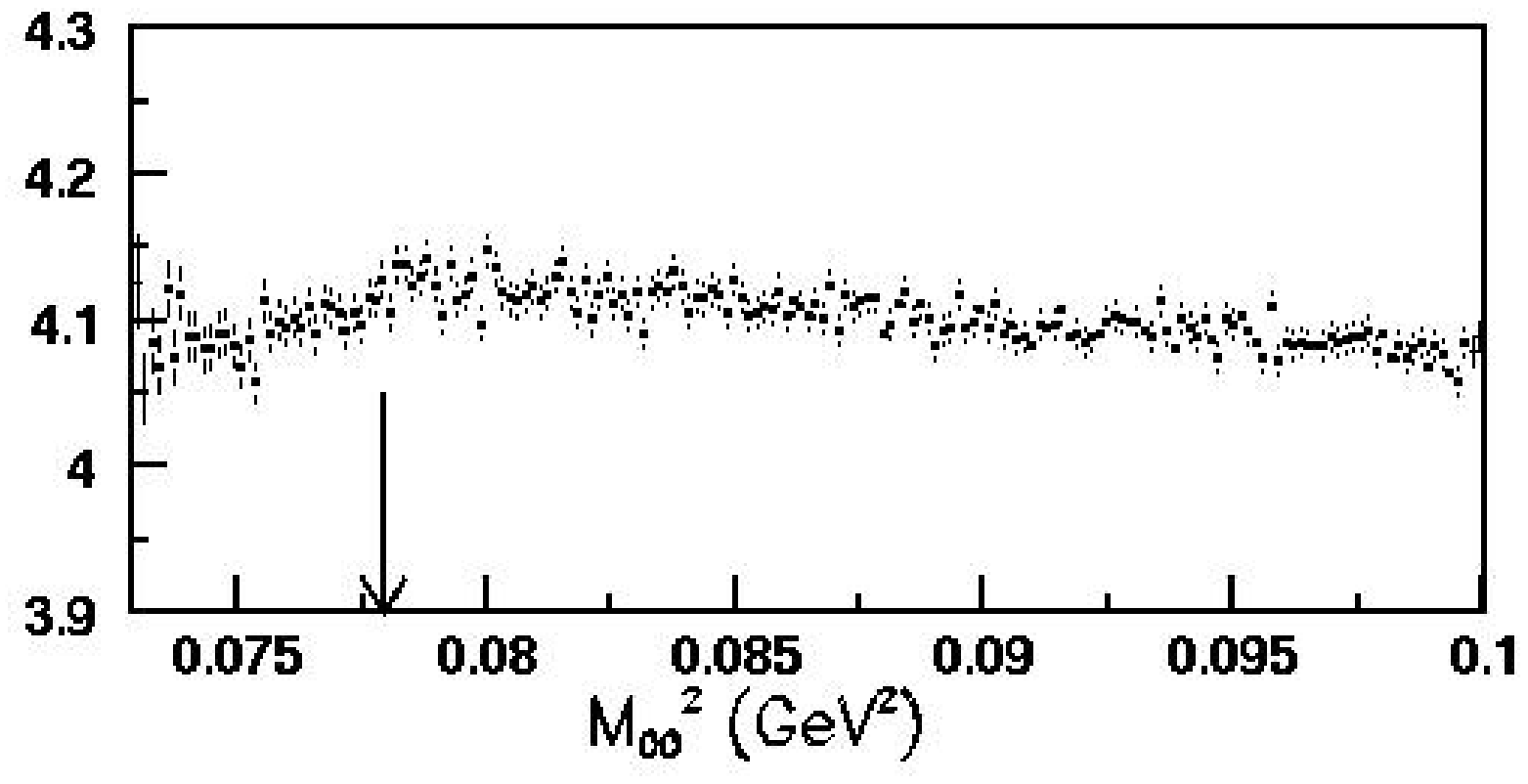}
    \caption{Distribution of $m^2_{\pi^0\pi^0}$ of the NA48 $\klpipipi$ data.
             Left: Full region. Right: Data divided by Monte Carlo simulated events
             without rescattering (arbitrary normalisation). The arrow marks
             the $(2 \, m_{\pi^+})^2$ threshold.} 
    \label{fig:klcusp}
  \end{center}
\end{figure}

The NA48 collaboration has analyzed about 88~million $\klpipipi$ 
events, recorded in a special data-taking period in 2000 with no 
material in front of the electro-magnetic calorimeter.
The data are practically back\-ground-free.
The distribution of squared $\pi^0\pi^0$ invariant mass is shown in
Fig.~\ref{fig:klcusp} (left).
It should be noted, that in the region of low invariant mass 
around the $\pi^+ \pi^-$ threshold the plot contains at maximum one entry per event,
and therefore no double-counting occurs.
When dividing the data with events from a 
Monte Carlo simulation without $\pi\pi$ rescattering, 
clear evidence for a change of slope near the cusp point is observed (Fig.~\ref{fig:klcusp}~(right)).

\section{Rescattering in $\kpmpipipi$ decays}

Effects from $\pi \pi$ rescattering should also occur in decays to charged pions
as in $\kpmpipipi$. In this case, $\pi^+ \pi^- \to \pi^+ \pi^-$ and
$\pi^\pm \pi^\pm \to \pi^\pm \pi^\pm$ contribute.
Here no cusp is possible inside the physical region, 
but the Dalitz plot distribution is modified.
The $K\to 3\pi$ matrix element, as a function of the Dalitz plot variables
$u = (s_3-s_0)/m^2_{\pi^+}$ and $v = (s_1-s_2)/m^2_{\pi^+}$, is given by~\cite{bib:pdg06}
\begin{equation}
  |{\cal M}|^2 = 1 + g \, u + h \, u^2 + k \, v^2 \, + \cdots,
\end{equation}
with the slope parameters $g$, $h$, and $k$.

NA48/2 has performed a combined fit to their $\kpmpipizpiz$ and $\kpmpipipi$ data, using
the non-relativistic effective field theory of the Bern-Bonn group~\cite{bib:bernbonn}.
Free parameters in the fit were the normalisations, $a_0 - a_2$, and the slope parameters
$g$, $h$, and $k$ of $\kpmpipipi$ and $g_0$ and $h_0$ of $\kpmpipizpiz$. The fit had a
nearly perfect $\chi^2$ of $757.1/757$~d.o.f. and resulted in significantly shifted values of the
slope parameters $g$, $h$, and $k$ with respect to a fit with no rescattering taken into account.
However, in the latter fit the $\chi^2$ value was equally good ($516.9/517$).
This means, that on the one hand $\pi\pi$ rescattering has a significant impact on the values of the Dalitz plot slopes.
On the other hand, there is practically no possibility to extract the scattering lengths from the $\kpmpipipi$
Dalitz plot slopes alone.

\section{Measurement of $\pi \pi$ Scattering Lengths in \\ $K_{e4}$ Decays}

An independent and complementary method for the determination of 
$\pi \pi$ scattering lengths is the analysis of $\kpmpipie$ ($K_{e4}$) decays.
The $K_{e4}$ decay is a rare decay with a branching fraction of about
$4 \times 10^{-5}$~\cite{bib:pdg06}.
Its amplitude depends on the two complex phases $\delta_0$ and
$\delta_1$, which are the $S$ and $P$ wave
$\pi \pi$ scattering phase shifts for isospin $I=0$.
In $K_{e4}$ decays, their difference $\delta = \delta_0 - \delta_1$ can be measured
as a function of $m_{\pi\pi}$.

\begin{figure}[t]
  \begin{center}
    \includegraphics[width=\linewidth]{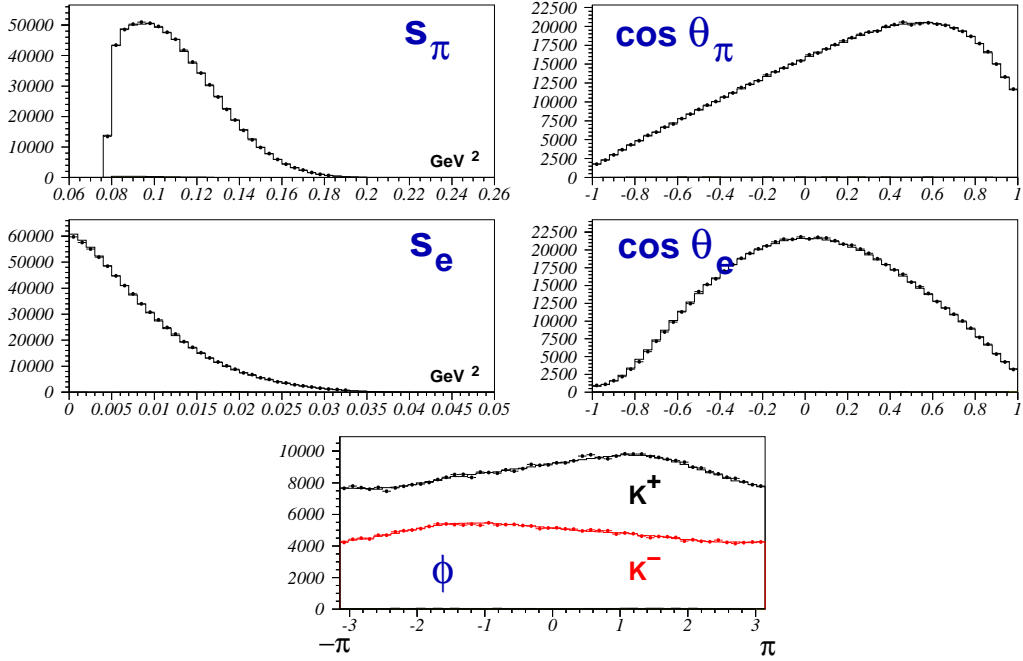}
    \caption{Distributions of the five Cabibbo-Maksymowicz variables in the
             NA48/2 2003 $K_{e4}$ data.}
    \label{fig:ke4cmplots}
  \end{center}
\end{figure}

In the 2003 data-taking period, 
the NA48/2 collaboration has collected $677510$ $K_{e4}$ decays 
with very small background contamination of only about $0.5\%$~\cite{bib:na48ke4}.
An analysis was carried out in the five independent Cabibbo-Maksymo\-wicz variables:
these are the squared invariant dipion and dilepton masses $s_\pi$ and $s_e$,
the angles $\theta_\pi$ and $\theta_e$ of $\pi^+$ and $e^+$ with respect to
the $\pi\pi$ and $e\nu$ directions in the $\pi\pi$ and $e\nu$ rest frames, 
respectively, and the angle $\phi$ between the $\pi\pi$ and $e\nu$ decay planes.
The distributions of these variables are shown in Fig.~\ref{fig:ke4cmplots}.

\begin{figure}[t]
  \begin{center}
    \includegraphics[width=0.7\linewidth]{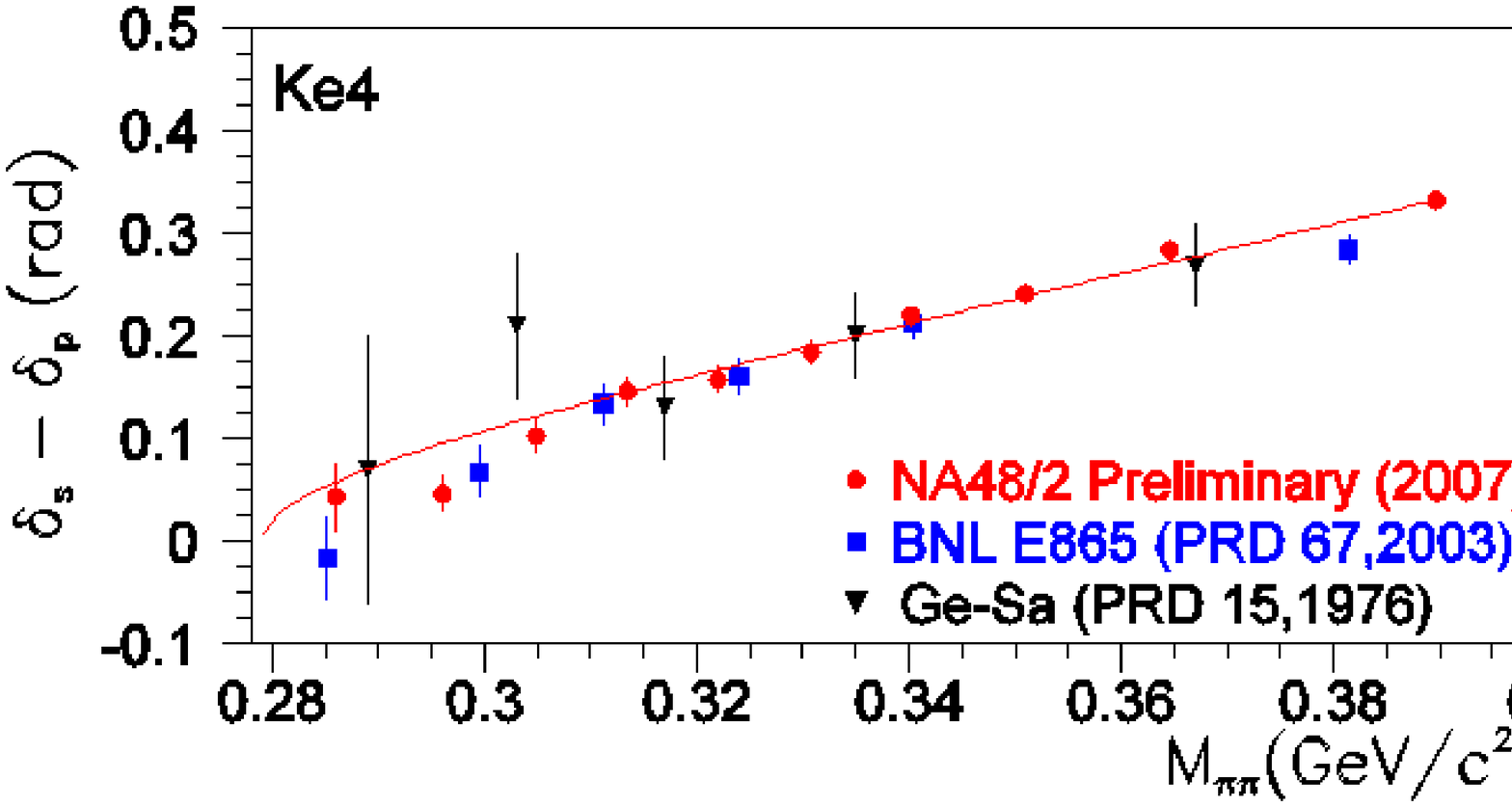}
    \caption{Measurements of the phase difference $\delta_s - \delta_p$
             as function of $m_{\pi\pi}$ in $K_{e4}$ decays. The line
             shows the best fit to $a_0$ and $a_2$ of NA48/2.}
    \label{fig:ke4results}
  \end{center}
\end{figure}

NA48/2 has performed a combined fit to the decay form factors and the
phase shift difference $\delta$ as a function of $m_{\pi\pi}$.
The results are shown in Fig.~\ref{fig:ke4results} together with the earlier
measurements of the Geneva-Saclay~\cite{bib:gesaclay}
and BNL E865~\cite{bib:e865} experiments. The results are in good agreement with each other 
(except perhaps the highest data point of E865). 
From the phase shift measurements, the $\pi\pi$ scattering lengths can be extracted.
Using dispersion relations (Roy equations) and data above 800~MeV, $a_2$ is 
related to $a_0$ (the so-called Universal Band). From a one-parameter fit, NA48/2 obtained
\begin{equation}
a_0 \, m_{\pi^+}= 0.256 \pm 0.006_\text{stat} \pm 0.002_\text{syst} {+0.018 \atop -0.017}_\text{theo},
\end{equation}
where the theoretical error corresponds to the width of the universal band, which is
given by the experimental uncertainties of the measurements above 800~MeV.
This result implies $a_2 \, m_{\pi^+}= -0.031 \pm 0.001_\text{exp} \pm {+0.013 \atop -0.012}_\text{theo}$.
When both $a_0$ and $a_2$ are left free in the fit, the results are
\begin{eqnarray}
\nonumber  a_0 \, m_{\pi^+} & = &  0.233 \pm 0.016_\text{stat} \pm 0.007_\text{syst}, \\
           a_2 \, m_{\pi^+} & = & -0.047 \pm 0.011_\text{stat} \pm 0.004_\text{syst}.
\end{eqnarray}

It has recently been realized, that effects from isospin symmetry breaking ($m_d \neq m_u$)
are not negligible and do have a significant impact on the value of the phase shift difference $\delta$. A preliminary ChPT calculation
gives a shift of $\delta$ between 12 and 15~mrad for $m_{\pi\pi}>300$~MeV/$c^2$~\cite{bib:ke4isobreaking}.
Taking into account these corrections, the value of $a_0$ shifts by $\sim -0.02$.
Note, that these corrections also have to be applied on the data of the previous $K_{e4}$ measurements.

\section{Conclusions}

In the past year, there has been significant progress in the extraction of
$\pi \pi$ scattering lengths from kaon decays. The NA48/2 experiment has discovered
a Wigner-cusp in the $m_{\pi^0\pi^0}$ distribution of $\kpmpipizpiz$ decays, arising
from $\pi \pi$ rescattering in this channel and thus allowing the precise extraction
of the scattering lengths $a_0$ and $a_2$. At the moment, the 
uncertainty is dominated by the neglect of radiative corrections and higher order terms.
Given the on-going theoretical work, this is expected to improve in the near future.
In addition, evidence for $\pi\pi$ rescattering in $\klpipipi$ events from
$\klpipipic$ has been found, using NA48 data from a $K_L$ run period.
The more traditional method for determination of $a_0$ and $a_2$ is the angular 
analysis of $K_{e4}$ decays, where the NA48/2 collaboration has presented
a new analysis with much larger precision than previous experiments.

\begin{figure}[t]
  \begin{center}
    \hspace{2mm}
    \includegraphics[width=0.44\linewidth]{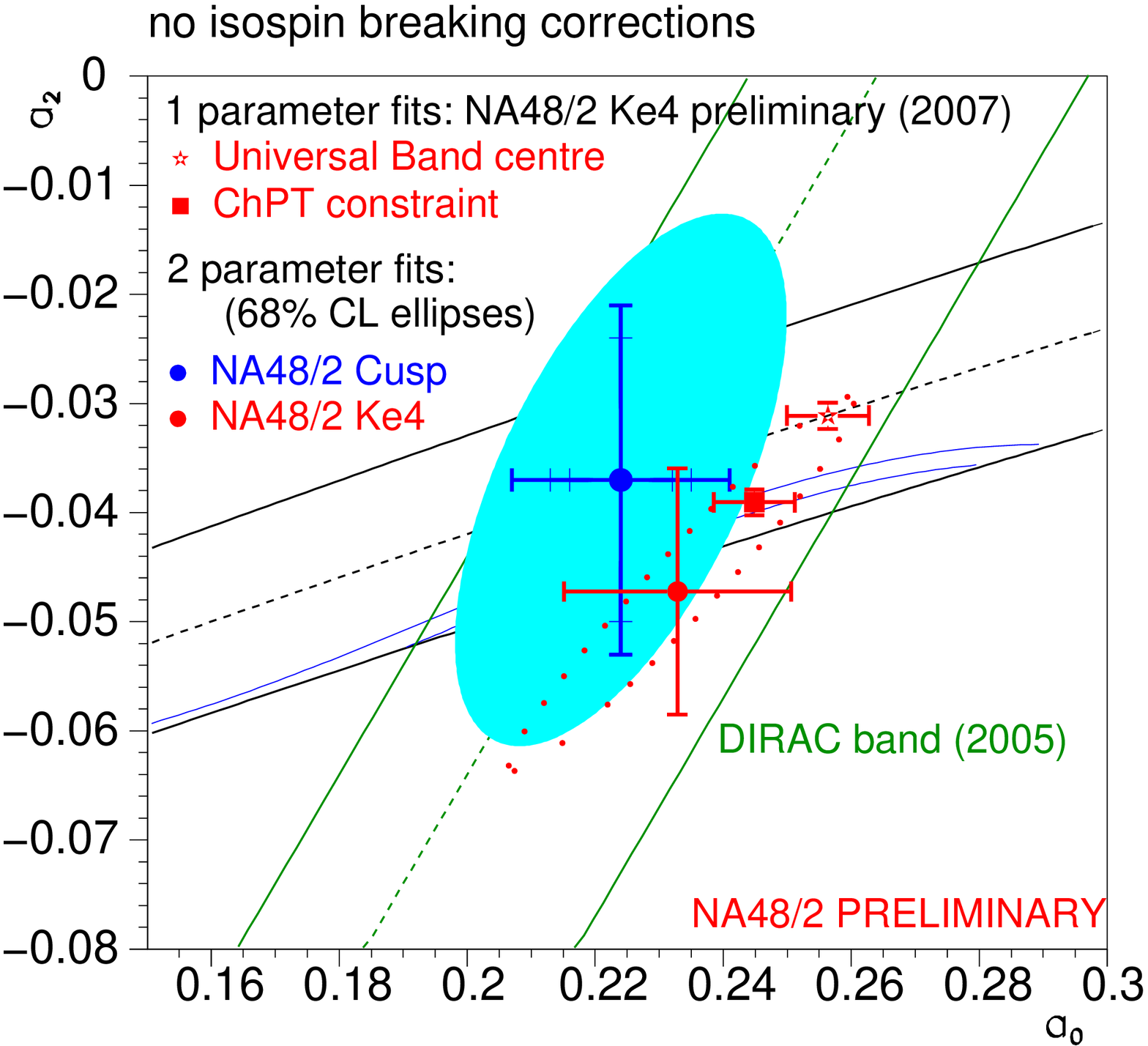}
    \hspace{\fill}
    \includegraphics[width=0.44\linewidth]{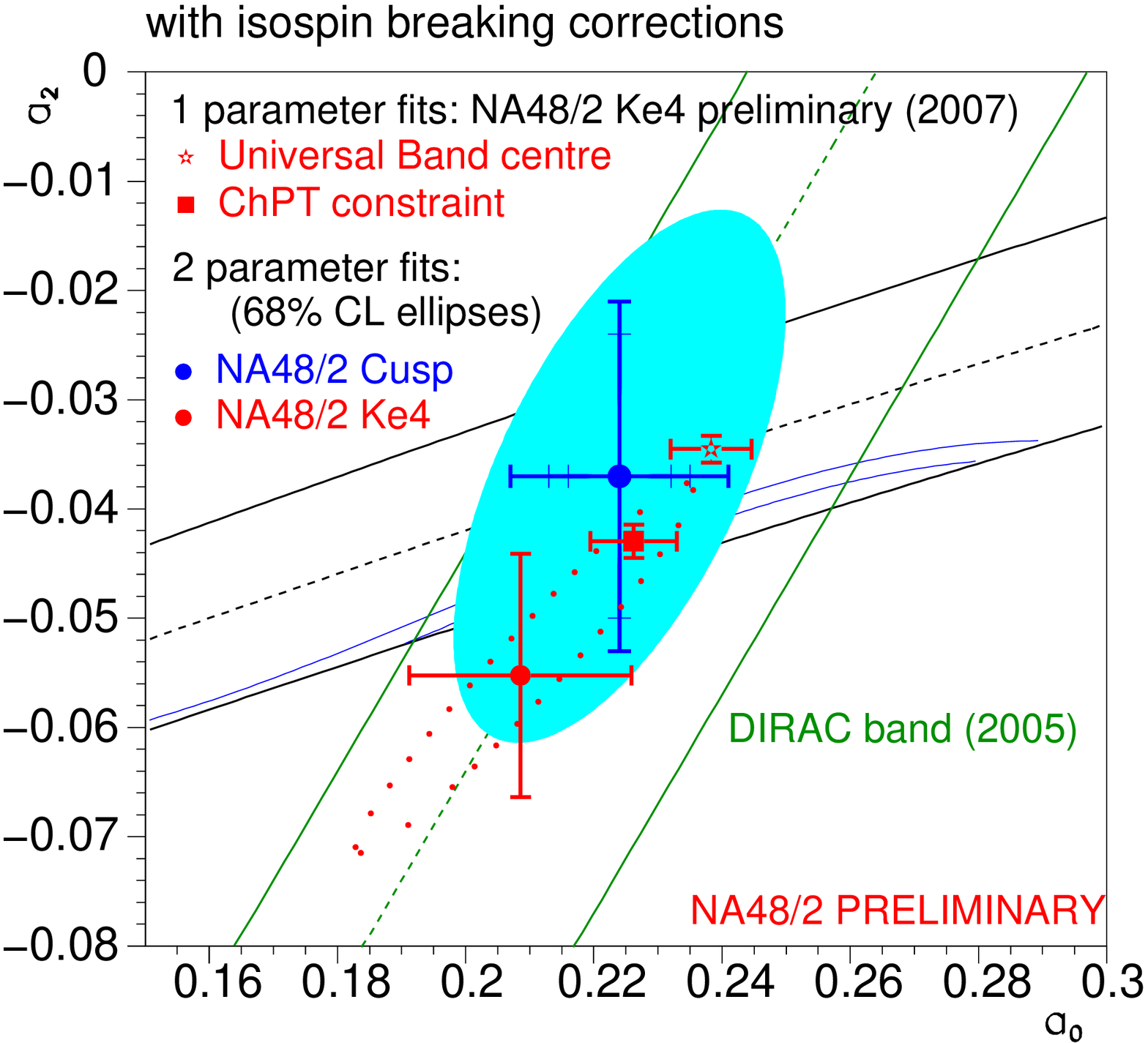}
    \hspace{2mm}
    \caption{Comparison of results on the scattering lengths $a_0$ and $a_2$ from the
             NA48/2 and DIRAC experiments without (left) and with (right)
             isospin breaking corrections applied to $K_{e4}$.
             Shown are also the universal band (black band), predicted using the 
             Roy equations and higher energy data, and the ChPT prediction
             (narrow blue band).}          
    \label{fig:comparison}
  \end{center}
\end{figure}

The constraints on $a_0$ and $a_2$ from the NA48/2 measurements,
together with the recent $a_0 - a_2$ determination
from pionium lifetime by DIRAC~\cite{bib:dirac}, are shown in
Fig.~\ref{fig:comparison}. When taking into account the corrections for isospin symmetry
breaking in $K_{e4}$, there is very good
agreement within the results and with the ChPT prediction. Further improvements
are expected in the near future from the analysis
of the complete NA48/2 $K_{e4}$ and DIRAC data sets, as well as from a better theoretical understanding
on the extraction of scattering lengths from $\kpmpipizpiz$ and $K_{e4}$.

\begin{thebibliography}{00}  



\bibitem{bib:gasserleutwyler} J.~Gasser and H.~Leutwyler,
                       \Journal{Annals~Phys.}{158}{142}{1984}.

\bibitem{bib:cgl}      G.~Colangelo, J.~Gasser, H.~Leutwyler,
                       \Journal{\NPB}{603}{125}{2001}.
                  

\bibitem{bib:gesaclay} L.~Rosselet {\it et al}. [Geneva-Saclay Collab.],
                       \Journal{\PRD}{15}{574}{1977}.


\bibitem{bib:e865}     S.~Pislak {\it et al}. [E865 Collab.],
                       \Journal{\PRL}{87}{221801}{2001};
                       \Journal{\PRD}{67}{072004}{2003}.

                        
\bibitem{bib:dirac}    B.~Adeva {\it et al}. [DIRAC Collab.],
                       \Journal{\PLB}{619}{50}{2005}.


\bibitem{bib:na48det}  V.~Fanti {\it et al.} [NA48 Collab.]
                       \Journal{\NIMA}{574}{433}{2007}.

\bibitem{bib:cabibbo}  N.~Cabibbo,
                       \Journal{\PRL}{93}{12181}{2004}.

\bibitem{bib:meissner}  U.-G..~Mei{\ss}ner, G.~M\"uller, S.~Steininger, 
                       \Journal{\PLB}{406}{154}{1997}; Erratum \Journal{\it ibid.}{407}{454}{1997}.

\bibitem{bib:cabibboisidori}  N.~Cabibbo, G.~Isidori,
                              \Journal{JHEP}{0503}{021}{2005}.

\bibitem{bib:na48cusp}    E.~Goudzovski,
                          to appear in {\it Proc.~KAON07~Conference}, Frascati, May 2007 (hep-ph/0706.4059).

\bibitem{bib:na48cuspold} J.R.~Batley {\it et al}. [NA48/2 Collab.],
                          \Journal{\PLB}{633}{173}{2006}.

\bibitem{bib:bissegger}   M.~Bissegger, A.~Fuhrer, J.~Gasser, B.~Kubis, and A.~Rusetsky, \\
                          \Journal{\PLB}{659}{576}{2008}.

\bibitem{bib:pdg06}    W.-M.~Yao {\it et al}. [Particle Data Group],
                       \Journal{\JPG}{33}{1}{2006}.

\bibitem{bib:cgl2} G.~Colangelo, J.~Gasser, H.~Leutwyler,
                   \Journal{\PRL}{86}{5008}{2001}.

\bibitem{bib:silagadze}  Z.K.~Silagadze,
                         \Journal{JETP~Lett.}{60}{689}{1994}.

\bibitem{bib:gevorkian}  S.R.~Gevorkian, A.V.~Tarasov, O.O.~Voskresenskaya,
                         \Journal{\PLB}{649}{159}{2007}.

\bibitem{bib:bernbonn}   G.~Colangelo, J.~Gasser, B.~Kubis, A.~Rusetsky,
                         \Journal{\PLB}{638}{187}{2006}.

\bibitem{bib:na48ke4}  J.R.~Batley {\it et al}. [NA48/2 Collab.],
                       CERN-PH-EP/2007-035, submitted to \EPJ.

\bibitem{bib:ke4isobreaking} J.~Gasser,
                             to appear in {\it Proc.~KAON07~Conference}, Frascati, May 2007 (hep-ph/0710.3048);
                             to appear in {\it Proc.~MENU~2007~Conference}, J\"ulich, September 2007.
                        

\end{thebibliography}
\end{document}